\documentstyle[12pt]{article}
\newcommand{\sect}[1]{\setcounter{equation}{0}\section{#1}}
 \begin{document}
\def\bq{\begin{equation}}
\def\eq{\end{equation}}
\begin{flushright}
 { 3 May }\\
{LPTENS-96/23 }\\
{UT-Komaba-96/11 }
\end{flushright}

\begin{center}
{\large\bf
CORRELATIONS OF NEARBY LEVELS INDUCED BY A RANDOM POTENTIAL } \end{center}

\begin{center}
{\bf E. Br\'ezin$^{a)}$ and S. Hikami$^{b)}$} \end{center}
\vskip 2mm
\begin{center}{$^{a)}$ Laboratoire de Physique Th\'eorique, Ecole Normale
Sup\'erieure}\\ {24 rue Lhomond 75231, Paris Cedex 05, France{\footnote{
Unit\'e propre du centre national de la Recherche Scientifique, Associ\'ee
\`a l'Ecole Normale Sup\'erieure et \`a l'Universit\'e de Paris-Sud} } }\\
{$^{b)}$ Department of Pure and Applied Sciences, University of Tokyo}\\
{Meguro-ku, Komaba, Tokyo 153, Japan}\\
\end{center}
\vskip 3mm
\begin{abstract}
We consider a Hamiltonian $H$ which is the sum of a deterministic part
$H_0$ and of a random potential $V$. For finite $N \times N$ matrices,
following a method introduced by Kazakov, we derive a representation of the
correlation functions in terms of contour integrals over a finite number of
variables. This allows one to analyse the level correlations, whereas the
standard methods of random matrix theory, such as the method of orthogonal
polynomials, are not available for such cases. At short distance we
recover, for an arbitrary $H_0$, an oscillating behavior for the connected
two-level correlation.
\end{abstract}
\newpage
\sect{Introduction}
\vskip 5mm
Let us first recall the results for the correlations between two
eigenvalues for the simple unitary ensemble, in which the full Hamiltonian
is treated as random. In the simplest Gaussian ensemble one considers $N
\times N$ random Hermitian matrices H with probability distribution
\begin{equation}\label{1.1}
P(H) = {1\over{Z}} {\rm exp} ( - {N\over{2}} {\rm Tr} H^2 ) \end{equation}
Two kinds of universal correlations between eigenvalues are known to be
present for such problems: a) a short-distance universal oscillatory
behavior; b) a finite distance universality of smoothed correlations.

Let
us review these two properties. The density of eigenvalues and the
two-level correlation function are defined as \begin{equation}\label{1.2}
\rho(\lambda) = < {1\over{N}}{\rm Tr} \delta ( \lambda - H) >
\end{equation} and
\begin{equation}\label{1.3}
\rho^{(2)}(\lambda,\mu) = < {1\over{N}}{\rm Tr} \delta ( \lambda - H)
{1\over{N}} {\rm Tr} \delta ( \mu - H ) > \end{equation}

The correlation function, when $\lambda$ and $\mu$ are arbitrary, has a
complicated, non-universal, oscillatory behavior. It simplifies and is
independent of the probability distribution of H, when

a) $\lambda- \mu$ is small, N is large, and the scaling variable
\begin{equation}\label{1.4}
x = \pi N ( \lambda - \mu ) \rho ({1\over{2}}(\lambda + \mu )) \end{equation}
is held finite. Then one finds\cite{Mehta} \begin{eqnarray}\label{1.5}
\rho^{(2)}_c(\lambda,\mu) &\simeq& {1\over{N}}
\delta(\lambda - \mu)\rho(\lambda) -
\rho(\lambda)\rho(\mu)
{\sin^2 x\over{x^2}}\nonumber\\
&\simeq& {1\over{N}} \delta(\lambda - \mu)\rho(\lambda) - {1\over{ \pi^2
N^2}}{\sin^2 [\pi N (\lambda - \mu) \rho ({\lambda + \mu \over{2}})]\over{(
\lambda - \mu )^2}}
\end{eqnarray}

b) Away from this short-distance region, for arbitrary $\lambda$ and $\mu$,
the correlations simplify only if one "smooths" the oscillations. This is
what one usually does, if one lets N go to infinity first in the resolvent,
before returning to the real axis. The result, which is known to be
universal, is \cite{BZ1}
\begin{equation}\label{1.6}
\rho^{(2)}_c(\lambda,\mu) = - {1\over{2 N^2 \pi^2}} {1\over{(\lambda - \mu
)^2}} {(a^2 - \lambda \mu)\over{[(a^2 - \lambda^2)(a^2 - \mu^2)]^{1/2}}}
\end{equation}
where $a$ is an end point of the support.

There are many equivalent derivations of the property b). They are based
either on orthogonal polynomials\cite{BZ1}, or on summing over planar
diagrams \cite{BZ2,BHZ1}, or solving an integral equation
\cite{Beenakker,Ambjorn};
however the property a) is known only through the orthogonal polynomials
approach \cite{BZ1}. For the generalization that we have in view in this
article, in which the "unperturbed" part of the Hamiltonian is
deterministic, if again for b) a diagrammatic approach still works
\cite{BZ2,BHZ1,BZ3,Zee}, we are not aware of any method which would allow
us to study whether a) still holds. To this effect we shall generalize a
method, introduced by Kazakov\cite{Kazakov}, to the study of correlation
functions. It consists of introducing an external matrix source. It leads
to a representation of the correlation function in terms of contour
integrals over two variables for finite N. We have used already this
representation in a previous paper devoted to random matrices made of
complex blocks, i. e. the Laguerre ensemble\cite{BHZ2}, but there we have
let the source go to zero at the end of the calculation. Keeping this
source finite allows one to deal with an arbitrary deterministic $H_0$. We
shall illustrate it here when the random potential V belongs to the simple
Gaussian unitary ensemble.

\vskip 5mm
\sect{ An external matrix source: deterministic plus random hamiltonian}
We consider a Hamiltonian $H = H_0 + V$, where $H_0$ is deterministic and
$V$ is a random $N \times N$ matrix. The Gaussian distribution P is given
by
\begin{eqnarray}\label{2.1}
P(H) &=& {1\over{Z}}e^{ - {N\over{2}} {\rm Tr} V^2 }\nonumber\\
&=&
{1\over{Z}}e^{- {N\over{2}}{\rm Tr} ( H^2 - 2 H_0 H + H_0^2)} \end{eqnarray}

We are thus simply dealing with a Gaussian unitary ensemble modified by a
matrix source $A= -H_0$. Up to a factor the probability distribution for H
is thus
\begin{equation}\label{2.2}
P_A(H) = {1\over{Z_A}}{\rm exp}( - {N\over{2}}{\rm Tr}H^2 - N {\rm Tr} A H
) \end{equation}
Let us first show how one deals with the density of states $\rho(\lambda)$
. It is the Fourier transform of the average "evolution" operator
\begin{equation}\label{2.3}
U_A(t) = < {1\over{N}}{\rm Tr} e^{i t H} > \end{equation}
and $\rho(\lambda)$ is
\begin{equation}\label{2.4}
\rho({\lambda}) =
\int_{-\infty}^{+\infty} {dt\over{2 \pi}} e^{- i t \lambda} U_A(t)
\end{equation}
We integrate first over the unitary matrix $\omega$ which diagonalizes H,
and without loss of generality we may assume that A is a diagonal matrix
with eigenvalues $(a_1, \cdots, a_N)$ . This is done by the well-known
Itzykson-Zuber integral \cite{Itzykson}, \begin{equation}\label{2.5}
\int
d{\omega} {\rm exp}( {\rm Tr} A {\omega} B {\omega} ^{\dag} ) =
{{\rm
det}({\rm exp}(a_i b_j))\over{\Delta(A)\Delta(B)}} \end{equation}
where
$\Delta(A)$ is the Van der Monde determinant constructed with the
eigenvalues of A:
\begin{equation}\label{2.6}
\Delta(A) = \prod_{i<j}^N (a_i - a_j)
\end{equation}
We are then led to
\begin{eqnarray}\label{2.7}
U_A(t) = && {1\over{Z_A \Delta(A)}}{1\over{N}}\sum_{\alpha =
1}^{N} \int
dr_1 \cdots dr_N e^{i t r_\alpha} \Delta(r_1,\cdots,r_N) \nonumber\\
&\times& {\rm exp}( - {N\over{2}}\sum r_i^2 - N \sum a_i r_i )
\end{eqnarray}
The normalization is
\bq\label{2.8}
U_A(0) = 1
\eq
The integration over the $r_i$ may be done easily, if we note that
\begin{eqnarray}\label{2.9}
&\int& dr_1 \cdots dr_N \Delta(r_1,\cdots,r_N) {\rm exp}( - {N\over{2}}\sum
r_i^2 - N \sum b_i r_i ) \nonumber\\
&=&
\Delta(b_1,\cdots,b_N) e^{{N\over{2}}\sum b_i^2} \end{eqnarray}
If we use
this, with
\begin{equation}\label{2.10}
b_i = a_i - {i t\over{N}} \delta _{\alpha,i} \end{equation} we obtain
\begin{equation}\label{2.11}
U_A(t) ={1\over{N}} \sum_{\alpha=1}^{N} \prod_{\gamma\neq \alpha}
({a_\alpha - a_\gamma - {i t \over{N}}\over{a_\alpha - a_\gamma}}) e^{-
{t^2\over{2N}} - i t a_\alpha }
\end{equation}
The sum over N terms may be replaced by a contour-integral in the complex u
plane,
\begin{equation}\label{2.12}
U_A(t) = - {1\over{i t}} \oint {du\over{2 \pi i}} \prod_{\gamma = 1}^{N}
({u - a_\gamma - {i t\over{N}}\over{ u - a_\gamma}}) e^{- i t u -
{t^2\over{2 N}} }
\end{equation}
The contour of integration encloses all the eigenvalues $a_\gamma$. If we
let all the $a_\gamma$ go to zero, we obtain \begin{equation}\label{2.13}
U_0(t) =
- {1\over{i t}}e^{- {t^2\over{2 N}}} \oint {du\over{2 \pi i}} e^{ - i t u }
( 1 - { i t \over{ N u}})^N
\end{equation}
From this exact representation for finite N, it is immediate to recover
all the well-known properties, the semi-circle law, or the more subtle edge
behavior of the density of states. Since the result is similar to the
Laguerre case that we have discussed in an earlier paper \cite{BHZ2}, we
shall not discuss it here.

For the two-level correlation function,
$\rho^{(2)}(\lambda,\mu)$ is
obtained from the Fourier transform $U_A(t_1,t_2)$, \begin{equation}\label{2.14}
\rho^{(2)}(\lambda,\mu) =
\int\int {dt_1 dt_2\over{(2 \pi)^2}} e^{- i t_1 \lambda - i t_2
\mu}U_A(t_1,t_2) \end{equation}
where $U_A(t_1,t_2)$ is
\begin{equation}\label{2.15}
U_A(t_1,t_2) = <{1\over{N}}{\rm Tr} e^{i t_1 H} {1\over{N}} {\rm Tr} e^{i
t_2 H} >
\end{equation}
The normalization conditions are
\begin{eqnarray}\label{2.16}
&& U_A(t_1,t_2) = U_A( t_2,t_1)\nonumber \\ && U_A(t_1,0) =
U_A(t_1)\nonumber\\ && U_A(0) = 1
\end{eqnarray}
Dealing with $U_A(t_1,t_2)$ is also simple; we have exposed the technique
in more details in the almost similar problem of the Laguerre
ensemble\cite{BHZ2}.
After performing the Itzykson-Zuber integral over the unitary group as in
(\ref{2.5}),
we obtain through the same procedure,
\bq\label{2.17}
U_A^{(2)}(t_1,t_2) =
{1\over{N^2}}\sum_{\alpha_1,\alpha_2 = 1}^N \int \prod_{i = 1}^N
dr_i {\Delta(r)\over{\Delta(A)}}e^{-N\sum ({1\over{2}}r_i^2 + r_i a_i) + i
(t_1 r_{\alpha_1} + t_2 r_{\alpha_2})} \eq After integration over the
$r_i$, we obtain
\begin{eqnarray}\label{2.18}
&& U_A(t_1,t_2) = {1\over{N^2}}\sum_{\alpha_1,\alpha_2} {\prod_{i<j}(a_i -
a_j - {i t_1\over{N}}(\delta_{i,\alpha_1} - \delta_{j,\alpha_1}) - {i
t_2\over{N}}(\delta_{i,\alpha_2}-
\delta_{j,\alpha_2}))\over{\prod_{i<j}(a_i - a_j)}}\nonumber\\ &&\times
e^{- i t_1 a_{\alpha_1} - i t_2 a_{\alpha_2} - {t_1^2\over{2N}} -
{t_2^2\over{2N}} - {t_1 t_2 \over{N}}\delta_{\alpha_1,\alpha_2}}
\end{eqnarray}
The terms of this
double sum in which $\alpha_1 = \alpha_2$ are written as a single contour
integral and their sum is simply ${1\over{N}}U_A(t_1 + t_2)$ of
(\ref{2.11}).
The Fourier transform of this term becomes \begin{equation}\label{2.19}
{1\over{N(2\pi)^2}}\int \int
dt_1 dt_2 e^{- i t_1 \lambda - i t_2 \mu} U_A (t_1 + t_2) =
{1\over{N}}
\delta ( \lambda - \mu) \rho (\lambda)
\end{equation}
The remaining part,
after the subtraction of the disconnected part, becomes
\begin{eqnarray}\label{2.20}
U_A(t_1,t_2) &&= -{1\over{N^2}}
\oint {du dv\over{(2\pi i)^2}}
e^{- {t_1^2\over{ 2 N}} - {t_2^2\over{2N}} - i t_1 u - i t_2 v} {1 \over{(u
- v - {i t_1\over{N}}) (u - v + {it_2\over{N}})}} \nonumber\\
&&\times \prod_{\gamma = 1}^N
( 1 - { i t_1\over{ N( u - a_\gamma)}})( 1 - {i t_2 \over{N (v - a_\gamma
)}}) \end{eqnarray}
where the contours are taken around $u = a_\gamma$ and $v = a_\gamma$. If
we include also the contour-integration around the pole, $v = u - {i t_1
\over{N}}$, this gives precisely the term $U_A(t_1 + t_2)$ of (\ref{2.19}),
which contributes to the delta-function part. This coincidence had already
been noticed for the Laguerre ensemble \cite{BHZ2}. We are now in position
to study the various properties of this random matrix problem with an
arbitrary source A.

\vskip 5mm
\sect{ Large N limit of the density of states}

For arbitrary A, the density of state $\rho(\lambda)$ was first found, in
the large N limit, by Pastur\cite{Pastur}. The result may be easily
recovered by summing diagrams.
Indeed in the large N limit the leading
diagrams are planar,
and the one-particle Green function is the sum of "rainbow" diagrams. It
follows immediately that the self-energy is proportional to the Green
function itself in the large N limit \cite{BZ2}, and this leads at once to
Pastur's result.
From the contour-integral representation (\ref{2.12}), let us show first
how to recover
this result . The average resolvent $G(z)$ is written in terms of the
evolution operator as
\begin{eqnarray}\label{3.1}
G(z) &=& <{1\over{N}}{\rm Tr} {1\over{z-H}} >\nonumber\\ &=&
i\int_{0}^{+\infty} dt e^{-i
t z} U_A(t)
\end{eqnarray}
We substitute (\ref{2.12}) for $U_A(t)$ and replace the product \bq\label{3.2}
\prod_{\gamma = 1}^N\left ( 1 - {it\over{N(u - a_\gamma)}}\right ) = {\rm
exp} \sum_{\gamma = 1}^N {\rm Log}\left ( 1 - {i t \over{N(u -
a_\gamma)}}\right )
\eq
by its leading term in the large N limit, namely \bq
{\rm exp}( - {it\over{N}}\sum_{\gamma = 1}^N {1\over{ u - a_\gamma}})
\eq

If we define the density of states of the external matrix A
\begin{equation}\label{3.4}
\rho_0(a) = {1\over{N}}\sum_{\alpha = 1}^N {\delta} (a -
a_{\alpha})\end{equation}
we may write this expression as
\bq\label{3.5}
{\rm exp} ( - i t \int da {\rho_0(a)\over{u - a}} ) \eq
Note that the "unperturbed" resolvent
\bq
G_0(z) = <{1\over{N}}{\rm Tr} {1\over{z-H_0}} >\eq is related to $\rho_0$ by
\bq\label{3.6}
G_0(z) = \int da {\rho_0(a)\over{u + a}}\eq since $H_0 = -A$.
We obtain then easily
\bq\label{3.7}
{\partial G\over{\partial z}} =
\oint {du\over{2 \pi i}} {1\over{u+G_0(u) -z}} \eq

We have now to specify the contour of integration in the complex u-plane.
It surrounds all the eigenvalues of $H_0$ and we have to determine the
location of the zeroes of the denominator with respect to this contour. Let
us return to the discrete form for the equation \bq\label{3.8}
u + G_0(u) = z
\eq

i.e.
\bq\label{3.9}
u + {1\over{N}}\sum_{i= 1}^N {1\over{u-\epsilon_i }}= z
\eq
which posseses $(N+1)$ real or complex roots in the u-plane. For z real and
large, N of these roots are close to the $\epsilon_i $ and one, which will
be denoted $\hat u(z)$, goes to infinity with z as \bq\label{3.10}
\hat u(z) = z - {1\over{z}} + O\left({1\over{z^2}}\right)\eq

Therefore, for large z, the contour encloses all the roots of (\ref{3.9}) except
$\hat u(z)$. When z decreases the contour should not be crossed by any
other root of the equation, therefore it is defined by the requirement that
only one root remains at its exterior. Therefore it is easier to calculate
the integral (\ref{3.7}) by taking the residues of the singularities
outside of the contour, rather than the N poles enclosed by this contour.
There are two of them outside; one is $\hat u(z)$ and the other one is at
infinity (since for large u, $G_0(u)$ vanishes). Taking these two
singularities we obtain
\begin{eqnarray}\label{3.11}
{\partial G\over{\partial z}} &=&
1 - {1\over{ 1+ {dG_0\over{d\hat u(z)}}}} \nonumber\\ &=&
1 -{d\hat u(z)\over{dz}} \end{eqnarray}
The integration gives
\bq\label{3.12}
G(z) = z - \hat u(z) \eq
(there is no integration constant since $G(z)$ vanishes for z large; note
that it does behave as it should as $1\over{z}$ for z large).
This combined with (\ref{3.8}) gives Pastur's self-consistent relation
\bq\label{3.13}
G(z) = G_0( z-G(z))
\eq

\sect{Universal correlations }
In the integral
representation (\ref{2.20}) we may neglect the terms $ t^2/N$ in the large
N limit and replace the products as in (\ref{3.5}). This gives the large
N-limit of $U_A^{(2)}(t_1,t_2)$ as \begin{equation}\label{4.1}
U_A^{(2)}(t_1,t_2) = - {1\over{N^2}}
\oint {du dv\over{(2\pi i)^2}}{1\over{(u - v)^2}} e^{-it_1 ( u + \int
{\rho_0(a)\over{u - a}}da ) - i t_2 ( v + \int {\rho_0(a)\over{ v - a}}da)}
\eq
Noting that
\bq\label{4.2}
{\partial^2\over{\partial z_1 \partial z_2}} {\rm ln} [ \hat u(z_1) - \hat
u(z_2) ] = {1\over{(\hat u(z_1) - \hat u(z_2))^2}} {d \hat u\over{d z_1}}
{d\hat u\over{d z_2}} \eq
we obtain,through identical steps, the connected two-particle Green
function \begin{eqnarray}\label{4.3}
G_c^{(2)}(z_1,z_2) &=& < {1\over{N}} {\rm tr} {1\over{z_1 - H}} {1\over{N}}
{\rm tr} {1\over{z_2 - H}} >_c\nonumber\\ &=& -
{1\over{N^2}}{\partial^2\over{\partial z_1 \partial z_2}} {\rm ln} [ \hat
u(z_1) - \hat u(z_2) ]
\end{eqnarray}
This result was derived earlier by diagrammatic methods \cite{BZ2}, and was
used to show that the singularity of the correlations, obtained when $z_1$
and $z_2$ approach the real axis with opposite imaginary parts, is
universal.

However if we want to study the correlation function in the short-distance
limit, we cannot use the resolvent any more (since we need to let the
imaginary parts of $z_1$, $z_2$ go to zero before N goes to infinity).

Returning then to (\ref{2.20}), and making the shifts, $t_1 \rightarrow t_1
- i u N$,
and $t_2 \rightarrow
t_2 - i v N$, the two-level
correlation function is remarkably  factorized since,
\begin{eqnarray}\label{4.5}
\rho_c(\lambda_1,\lambda_2) &=& \int {dt_1\over{2 \pi}} \oint {dv\over{2
\pi i}}\prod_{\gamma=1}^N
({a_\gamma + { i t_1\over{N}}\over{
v - a_\gamma}}) {1 \over{v + {i t_1\over{N}}}}e^{- {N\over{2}} v^2 -
{t_1^2\over{2N}} - i t_1 \lambda_1 - N v \lambda_2} \nonumber\\ &&\times
\int {dt_2\over{2 \pi}}\oint {du\over{2 \pi i}}
\prod_{\gamma=1}^N({a_\gamma +
{ i t_2\over{N}}\over{u - a_\gamma}})
{1\over{u + {i t_2\over{N}}}}e^{- {N\over{2}} u^2 - {t_2^2\over{2N}} - i
t_2 \lambda_2 - N u \lambda_1} \nonumber\\ && = - K_N(\lambda_1,\lambda_2)
K_N(\lambda_2,\lambda_1) \end{eqnarray} This kernel
$K_N(\lambda_1,\lambda_2)$ is further simplified by the shift $t_1
\rightarrow t + i v N$, \bq
K_N(\lambda_1,\lambda_2) =
\int {dt\over{2 \pi}} \oint {dv\over{2 \pi i}} {1\over{i t}}\prod_{\gamma=1}^N
( 1 - { it \over{
N( v - a_\gamma)}})
e^{ - {t^2\over{2N}} - i v t - i t \lambda_1 + N v (\lambda_1 - \lambda_2)} \eq
Note that $K_N(\lambda_1,\lambda_1)$ reduces to the density of states.
We replace again the product in (\ref{4.5}) by its large N-limit, neglect
${t^2\over{N}}$ and integrate over $t$, leading to \bq\label{4.6}
{\partial K_N\over{\partial \lambda_1}}
= {1\over{\pi}} {\rm Im} \oint {du\over{2\pi i}} {1\over{ u + G_0(u) -
\lambda_1 + i \epsilon}} e^{- u y} \eq
with $y = N(\lambda_1 - \lambda_2)$.
Therefore
\begin{eqnarray}\label{4.7}
{\partial K_N\over{\partial \lambda_1}} &=& {1\over{\pi}} {\rm Im} {d \hat
u\over{d \lambda_1}} e^{- y \hat u (\lambda_1 - i \epsilon)}\nonumber\\
&=& - {1\over{\pi y}} {\partial \over{\partial \lambda_1}} {\rm Im} \left (
e^{- y \hat u (\lambda_1 - i \epsilon)} \right )
\end{eqnarray}
Since, from (\ref{3.12}),
\bq\label{4.8}
\hat u (\lambda_1 - i \epsilon ) = \lambda_1 - {\rm Re} G(\lambda_1) - i
\pi \rho(\lambda_1)
\eq
we
obtain
\bq\label{4.9}
K_N(\lambda_1,\lambda_2) = - {1\over{\pi y}} e^{- y [ \lambda_1 - {\rm
Re}G(\lambda_1) ]} {\rm sin} [\pi y \rho(\lambda_1)]
\eq
Repeating this calculation for $K_N(\lambda_2,\lambda_1)$ we end up, in the
large N, finite $y$ limit,  with

\bq
\rho_c(\lambda_1,\lambda_2) =
-{1 \over{ \pi^2 y^2}} {\rm sin}^2 [ \pi \rho( {\lambda_1 +
\lambda_2\over{2}}) y ] \eq
Note that this result is independent of $H_0$ (apart from the scale factor
present in the densitty of states). In the case in which   $H_0$ vanishes
it is also independent of the probability distribution of V\cite{BZ1}. It
is natural to conjecture that this remains true for  $H_0$ non-zero as
well, but we do not know of any method to prove it.

\vskip 5mm
\sect{Conclusion }
The exact expressions for
finite N of the correlation functions in terms of contour integrals, have
allowed us to study the short-distance correlations for an arbitrary
unperturbed Hamiltonian.  We are not aware of any other method which could
be used for solving this problem. The result is, as expected, a universal
short distance behaviour, which depends on  $H_0$ only  through scale
factors.

\newpage
\begin{center}
{\bf Acknowledgement}
\end{center}
We are thankful for the support by the cooperative research project between
the Centre National de la Recherche Scientifique and the Japan Society for
the Promotion of Science.


\vskip 5mm
\begin{thebibliography}{99}
\bibitem{Mehta} M. L. Mehta, Random matrices, 2nd ed. (Academic Press, New
York 1991).
\bibitem{BZ1} E. Br\'ezin and A. Zee, Nucl. Phys. {\bf B 402} (1993) 613.
\bibitem{BZ2} E. Br\'ezin and A. Zee, Phys. Rev. {\bf E 49} (1994) 2588.
\bibitem{BHZ1} E. Br\'ezin, S. Hikami and A. Zee, Phys. Rev {\bf E 51}
(1995) 5442.
\bibitem{Beenakker} C. W. J. Beenakker, Nucl. Phys. {\bf B 422}, 515 (1994).
\bibitem{Ambjorn} J. Ambjorn and Yu. M. Makeenko, Mod. Phys. Lett. {\bf
A 5}, 1753 (1990).
\bibitem{BZ3} E. Br\'ezin and A. Zee, Nucl. Phys. {\bf B 453} (1995) 531.
\bibitem{Zee} A. Zee, a preprint
NSF-ITP-96-12, cond-mat/9602146.
\bibitem{Kazakov} V. A. Kazakov, Nucl. Phys. {\bf B 354} (1991) 614.
\bibitem{BHZ2} E. Br\'ezin, S. Hikami and A. Zee, Nucl. Phys. {\bf B
464} (1996) 411.
\bibitem{Itzykson} C. Itzykson and J. -B. Zuber, J. Math. Phys. {\bf 21}
(1980) 411.
\bibitem{Pastur} L. A. Pastur, Theor. Math. Phys. (USSR) {\bf 10}, 67
(1972). \end{thebibliography}
\end{document}